\documentclass{ws-jbcb}
\usepackage{amssymb,amsmath}

\begin{document}

\markboth{Srihari and Leong}
{A Survey of Methods for Complex Detection from Protein Interaction Networks}

\catchline{}{}{}{}{}
%
%%%%%%%%%%%%%%%%%%%%%%%%%%%%%%%%%%%%%%%%%%%%%%%%%%%%%%%%%%%%%%%%%%%%

\title{A SURVEY OF COMPUTATIONAL METHODS FOR\\ PROTEIN COMPLEX PREDICTION FROM\\ PROTEIN INTERACTION NETWORKS
%\footnote{For the title, try not to 
%use more than 3 lines. Typeset the title in 10~pt 
%Times roman, uppercase and boldface.} 
}

\author{SRIGANESH SRIHARI
%\footnote{Typeset names in 8~pt roman,  
%uppercase. Use the footnote to indicate the
%present or permanent address of the author.}
}

\address{Department of Computer Science, National University of Singapore, Singapore 117417\\
%\footnote{State completely without abbreviations, the
%affiliation and mailing address, including country. Typeset in 8~pt
%Times italic.}\\
srigsri@comp.nus.edu.sg}

\author{HON WAI LEONG}

\address{Department of Computer Science, National University of Singapore, Singapore 117417\\
leonghw@comp.nus.edu.sg}

\maketitle

%\begin{history}
%\received{(Day Month Year)}
%\revised{(Day Month Year)}
%\accepted{(Day Month Year)}
%\comby{(xxxxxxxxxx)}
%\end{history}

\begin{abstract}
Complexes of physically interacting proteins are one of the fundamental functional units responsible for driving
key biological mechanisms within the cell. 
Their identification is therefore necessary not only to understand complex formation but also the higher level
organization of the cell.
With the advent of ``high-throughput" techniques in molecular biology, significant amount of physical interaction data has been
catalogued from organisms such as yeast, which has in turn
fueled computational approaches to systematically mine complexes
from the network of physical interactions among proteins (PPI network). In this survey, we review, classify and evaluate some of the 
key computational methods developed till date for the identification of protein complexes from PPI networks. 
We present two insightful taxonomies that reflect how these methods have evolved over the years towards 
improving automated complex prediction.
We also discuss some open challenges facing accurate reconstruction of complexes, the crucial ones being
presence of high proportion of errors and noise in current high-throughput datasets and some key aspects overlooked by
current complex detection methods.
We hope this review will not only help to condense the history of computational complex detection for easy reference, but also
provide valuable insights to drive further research in this area.
\end{abstract}


\section{Introduction}
Most biological processes within the cell are carried out by proteins that physically \emph{interact} to form 
stoichiometrically stable \emph{complexes}.
Even in the relatively simple model organism {\em Saccharomyces cerevisiae} (budding yeast),
these complexes are comprised of many subunits that work in a coherent fashion.
These complexes interact with individual proteins or other complexes to form functional modules and pathways that drive
the cellular machinery. Therefore, a faithful reconstruction of the entire set of complexes (the `complexosome')
from the physical interactions among proteins (the `interactome') is essential to not only understand complex formations, 
but also the higher level cellular organization. 

Protein complexes constitute \emph{modular} functional units within the network of physical interactions, the
PPI network\cite{Spirin2003}.
From a biological perspective, this modularity is a result division of labor and of evolution to provide robustness against 
mutation and chemical attacks\cite{Hartwell1999}.
From a topological perspective, this modularity is a result of proteins within complexes being densely
connected to each other than to the rest of the network\cite{Zhang2008}.

Since the advent of ``high-throughput" screening in molecular biology in the late 1990s and early 2000s, 
several techniques have been introduced to infer physical interactions among proteins within organisms in a large-scale (``genome-wide") manner. 
These have helped to catalogue significant amount of protein
interactions in organisms such as yeast thereby fueling computational techniques to systematically mine and analyze such large-scale interaction data.
In yeast, the Yeast two-hybrid (Y2H)\cite{Uetz2000,Ito2001}, Protein Complementation Assay (PCA)\cite{Michinck2003} and
Tandem Affinity Purification followed by Mass Spectrometry (TAP-MS)\cite{Gavin2002,Ho2002,Gavin2006,Krogan2006} are some of the
widely adopted experimental systems that have helped to identify a considerable fraction of physical interactions among proteins. 
However, even at the current `state-of-the-art', these high-throughput techniques have been shown to produce
considerable proportion of spurious (false positive) interactions\cite{vonMering2002,Bader2002,Cusick2008,Mackay2008}.
Therefore, once the interactions are identified their qualities need to be first assessed to generate a reliable set of
interactions that is deemed suitable for further mining and analysis. This process includes assigning each interaction a confidence score 
that typically accounts for the biological variability and technical limitations of the experimental conditions, and therefore reflects
the reliability of the inferred interaction
\cite{Collins2007,Hart2007,Chua2008,Liu2008,Friedel2008,Kuchaiev2009,Voevodski2009,Jain2010,Suthram2006,SrihariThesis2012}.
The interactions with confidence scores below a certain threshold are discarded to build a reliable ``cleaned-up" PPI network.
This PPI network is then mined to identify groups of proteins potentially forming complexes. 
The whole process can be summarized in the following steps:
\begin{enumerate}
\item Integrating high-throughput datasets from multiple experiments and assessing the reliabilities of interactions;
\item Constructing a reliable PPI network;
\item Identifying modular subnetworks from the PPI network to generate a candidate list of complexes;
\item Evaluating the identified complexes against \emph{bona fide} complexes, and validating and assigning roles to
novel complexes.
\end{enumerate}

The identification of complexes from high-throughput interaction datasets has attracted considerable attention from both 
biologists as well as computational research communities, and over the years, several computational techniques have been developed to
systematically identify complexes. Quite naturally, a number of surveys have come out from time-to-time evaluating and comparing these techniques
for their performance on available PPI datasets.
One of the earliest comprehensive evaluations was by Brohee and van Helden (2006)\cite{Brohee2006}. This was followed by 
Vlashblom et al. (2009)\cite{Vlasblom2009} and Li et al. (2010)\cite{Xli2010}. While Brohee and Vlasblom et al. evaluated and 
compared some early methods on PPI datasets available at that time (till 2006), 
Li et al. covered some of the more recent methods developed until 2009.
The purpose of our work is to provide an up-to-date survey, classification (taxonomy) and evaluation of some the 
representative works done till date (2011/2012). We build upon the existing surveys so as to not repeat entirely the 
evaluations and conclusions already drawn, yet we provide our own classifications and evaluations of more recent techniques
across up-to-date PPI datasets. We also compare across unscored (raw) and scored PPI datasets, 
which is missing in these existing surveys. We also highlight and comment on some of the newer challenges and open problems in complex prediction,
which can guide future directions for research in this area.
%%%%%%%%%%%%%%%%%%%%

\section{Review of existing methods for complex detection}
We begin by mentioning some definitions and terminologies widely adopted across the reviewed works.
A PPI network is modeled as an undirected graph $G=(V,E)$, where $V$ is the set of proteins and $E = \{(u,v): u,v \in V\}$ is
the set of interactions among protein pairs.  For any protein $v \in V$, $N(v)$ is the set of direct neighbors of $v$,
while $deg(v) = |N(v)|$ is the degree of $v$. The interaction density of $G$ is
defined as $density(G) = \displaystyle \frac{2.|E|}{|V|.(|V|-1)}$. This is a real number between 0 and 1, and typically
quantifies the ``richness of interactions" within $G$: 0 for a network without any interactions 
and 1 for a fully connected network. The clustering coefficient $CC(v)$ measures the ``cliquishness"
of the neighborhood of $v$: $CC(v) = \displaystyle \frac{2.|E(v)|}{|N(v)|.(|N(v)|-1)}$, where $E(v)$ is
the set of edges in the neighborhood of $v$.
If the interactions of the network are reliability scored (weighted), these definitions 
can be extended to their corresponding weighted versions: 
$deg_w(v) = \displaystyle \sum_{u\in N(v)} w(u,v)$,
$density_w(G) = \displaystyle \frac{\displaystyle \sum_{e \in E}w(e)}{|V|.(|V|-1)}$, 
and $CC_w(v) = \displaystyle \frac{\displaystyle \sum_{e \in E(v)}w(e)}{|N(v)|.(|N(v)|-1)}$,
where $w: E \times E \rightarrow \mathcal{R}$ is a scoring function on the interactions in $E$.
There are several interesting variants proposed for weighted clustering coefficient $CC_w$;
for a survey see\cite{Kalna2007}.

\subsection{Taxonomy of existing methods}

Although at a very generic level most existing methods make the key assumption that complexes are embedded among
densely-interacting groups of proteins within PPI networks, these methods vary considerably either in the algorithmic
methodologies or the kind of biological insights employed to detect complexes.
Accordingly, we classified some popular complex detection methods into two broad categories (a soft classification):
(i) methods based solely on graph clustering;
(ii) methods based on graph clustering and some additional biological insights.
These biological insights may be in the form of
functional, structural, organizational or evolutionary information known about complexes 
or their constituent proteins from experimental or other biological studies.

\begin{figure}[th]
\centerline{\psfig{file=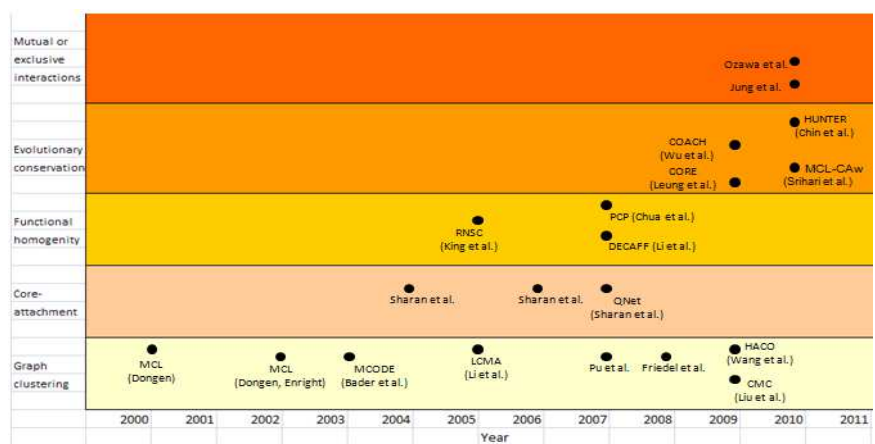,width=11.5cm,height=6.0cm}}
\vspace*{8pt}
\caption{The ``Bin-and-Stack" classification: 
Chronological binning of complex detection methods based on biological information used.
It is interesting to note that over the years, as researchers have tried to improve the basic graph clustering
ideas, they have also incorporated biological information into their methods.
}
\label{figure:Taxonomy_chronology}
\end{figure}

\begin{figure}[th]
\centerline{\psfig{file=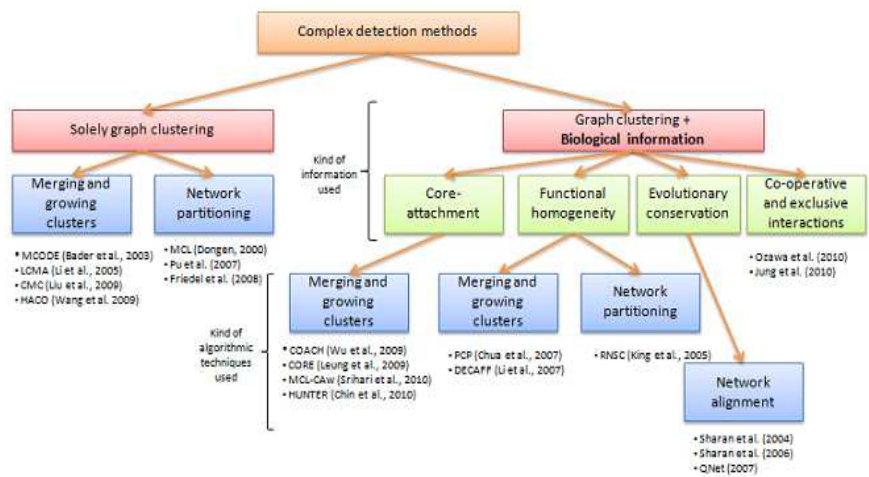,width=11.5cm,height=6.5cm}}
\vspace*{8pt}
\caption{The Tree classification: Classification of existing methods for complex detection based on the algorithmic 
methodologies used. Primarily three methodologies are adopted: merging and growing clusters, network partitioning and network alignment.}
\label{figure:Taxonomy_complex_detection_methods}
\end{figure}

We present this classification in two snapshots.
The first snapshot, shown in Figure~\ref{figure:Taxonomy_chronology}, gives a \emph{chronology-based} ``bin-and-stack"
classification, while the second snapshot, shown in Figure~\ref{figure:Taxonomy_complex_detection_methods}
gives a \emph{methodology-based} ``tree" classification of the methods.

In the chronology-based classification, we \emph{binned} methods based on the years in which they were developed,
and stacked them based on the kind of biological insights used (see Figure~\ref{figure:Taxonomy_chronology}).
The biological insights are grouped as: core-attachment structure, evolutionary information, functional coherence, and
mutually exclusive and co-operative interactions.
It is interesting to note from this classification that, over the years, as researchers tried to improve
the basic graph clustering ideas, they also incorporated a variety of biological information into their methods.

In the methodology-based classification, we distributed the methods to different branches of a
\emph{classification tree} based on the kind of computational strategy used (see Figure~\ref{figure:Taxonomy_complex_detection_methods}).
At the first level from the root, we grouped these methods into those based solely on graph clustering, and those
employing additional biological insights.
At subsequent levels, we further divided these methods based on the kind of algorithmic strategies used, into:
(i) methods employing merging or growing of clusters;
(ii) methods employing repeated partitioning of networks; and
(iii) methods employing network alignment.
The methods employing merging or growing clusters go ``bottom-up", that is, typically start with small ``seeds" 
(for example, triangles or cliques), and repeatedly add or remove proteins or merge clusters 
based on some similarity measures to arrive at the final set of complexes. 
On the other hand, the methods based on network partitioning go ``top-down", that is, repeatedly partition or break the network
into multiple subnetworks based on certain divisive criteria. The methods based on network alignment use multiple networks
(typically from different species) to arrive at isomorphic regions that likely correspond to complexes,
the inituition being that proteins belonging to real complexes should generally be conserved through the evolution
process to act as an integrated functional unit\cite{Zhang2008}.

\subsection{Methods based solely on graph clustering}
Most methods that cluster the PPI network into multiple dense subnetworks make use of solely the topology
of the network.

\subsubsection*{Molecular COmplex DEtection (MCODE)}
MCODE, proposed by Bader and Hogue (2003)\cite{Bader2003}, is one of the first computational methods (and therefore, seminal)
developed for complex detection from PPI networks.
The MCODE algorithm operates in mainly in two stages,
vertex weighting and complex prediction, and an optional third stage for post-processing.

In the first stage, each vertex $v$ in the network $G=(V,E)$ is weighted based on its neighborbood density. Instead of directly
using clustering coefficient, MCODE uses core-clustering coefficient which measures the density
of the highest $k$-core in the neighborhood of $v$. This amplifies the weighting of densely connected
regions in $G$. In the second stage, the vertex $v$ with the highest weight is used to seed a complex. MCODE then
recursively moves outwards from the seed vertex, including vertices into the complex whose weight is a 
given percentage (vertex weight parameter - VWP) away from the seed vertex. A vertex once added to a complex
is not checked subsequently. The process stops when there are no more vertices to be added to the complex,
and is repeated using the next unseeded vertex. At the end of this process multiple non-overlapping complexes are generated.
The optional third stage performs a post-processing on the complexes generated from the second stage. Complexes without
2-cores are filtered out, and new vertices in the neighborhood with weights 
higher than a given `fluff' parameter are added to existing complexes. The resultant complexes are scored and ranked 
based on their densities. The time complexity of the algorithm is $O(|V|.|E|.h^3)$, where $h$ is the vertex size 
of the average vertex neighbourhood in the network $G$.

\subsubsection*{Markov CLustering (MCL)}

The Markov Clustering (MCL) algorithm, proposed by Stijn van Dongen (2000)\cite{Dongen2000}, is a general graph
clustering algorithm that simulates random walks (called {\em flow}) to extract out relatively dense regions within
networks. In biological applications, it was first applied to cluster protein families and ortholog groups\cite{Enright2002} 
before it proved to be effective in detecting complexes from protein interaction networks\cite{Friedel2008,Pu2007,Enright2004}.

%\begin{figure*}[htp]
%\begin{center}
%\includegraphics[scale=0.60]{MCL_depiction.eps}
%\caption{How MCL works: Repeated expansion and inflation in MCL separates the network into multiple non-overlapping regions.}
%\label{figure:MCL_depiction}
%\end{center}
%\end{figure*}

MCL manipulates the adjacency matrix of networks with two operators called \emph{expansion} and \emph{inflation} to
control the random walks (flow). Expansion models the spreading out of the flow, while inflation models the
contraction of the flow, making it thicker in dense regions and thinner in sparse regions.
These parameters boost the probabilities of intra-cluster walks and demote those of inter-cluster walks. 
Mathematically, expansion coincides with normal matrix multiplication, while
inflation is a Hadamard power followed by a diagonal scaling. Therefore, MCL is highly efficient and scalable.
The iterative expansion and inflation separates the network into multiple
non-overlapping regions. 
%This is depicted in Figure~\ref{figure:MCL_depiction}; adopted from\cite{Dongen2000}.

\subsubsection*{Clustering based on merging Maximal Cliques (CMC)}
CMC was proposed by Liu et al. (2009)\cite{Liu2009} to detect complexes from PPI networks based
on repeated merging of maximal cliques. Some earlier algorithms like CFinder\cite{Adamcsek2003} and 
Local Clique Merging Algorithm (LCMA)\cite{Li2005} also adopted clique merging to find dense neighborhoods, 
but the distinct advantage of CMC over these algorithms
is its ability to work on weighted networks and to find relatively low density regions
(in subsequent improved versions of CMC).

CMC begins by enumerating all maximal cliques in the PPI network using the Cliques algorithm proposed
by Tomita et al.\cite{Tomita2006}. Although enumerating all maximal cliques is NP-hard, this does not
pose a problem in PPI networks because these networks are usually sparse. CMC then assigns a score
to each clique $C$ based on its weighted density, which considers the reliabilities (weights)
of the interactions within the clique:

\begin{equation}
Score(C) = \frac{\sum_{u,v \in C} w(u,v)} {|C|.(|C|-1)}.
\end{equation}

CMC then ranks these cliques in decreasing order of their scores and
iteratively merges or removes highly overlapping cliques based on their inter-connectivity scores. 
The inter-connectivity score of two cliques $C_i$ and $C_j$ is based on the non-overlapping regions of the two cliques
and is defined as:

\begin{equation}
Inter\_score(C_i,C_j) = 
\displaystyle \sqrt{
\displaystyle \frac{\sum_{u \in (C_i - C_j)} \sum_{v \in C_j} w(u,v)}{|C_i - C_j|.|C_j|}.
\displaystyle \frac{\sum_{u \in (C_j - C_i)} \sum_{v \in C_i} w(u,v)}{|C_j - C_i|.|C_i|}
}
\end{equation}

CMC determines whether two cliques $C_i$ and $C_j$ sufficiently overlap: $|C_i \cap C_j|/|C_j| \geq overlap\_thresh$.
If so, $C_j$ is either removed or merged with $C_i$ based on the inter\_score:
if the $inter\_score (C_i, C_j) \geq merge\_thresh$, then $C_i$ and $C_j$ are merged, else $C_j$ is removed.
Finally, all the resultant merged clusters are output as the predicted complexes.

\subsubsection*{Clustering with Overlapping Neighborhood Expansion (ClusterONE)}
Nepusz et al. (2012)\cite{Nepusz2012} proposed ClusterONE, a method for detecting overlapping protein complexes 
from weighted PPI networks, based on seeding and greedy growth, similar to MCODE\cite{Bader2003}. ClusterONE uses a cohesiveness measure to determine
how likely a group of proteins form a complex, and is based on the weight of the interactions within the group and with the rest of the network.

To begin with, ClusterONE identifies seed proteins and greedily grows them into groups with high cohesiveness. When the greedy growth for a group cannot progress
any more, a next seed protein is selected to repeat the procedure until no more seed proteins remain.
In the second step, ClusterONE identifies highly overlapping cohesive groups and merges them into potential complex candidates. ClusterONE allows identification of
overlapping complexes if each of the merged groups represent individual complexes that share proteins. Nepusz et al.'s comparisons with 
methods like MCODE, MCL and CMC showed that the complexes from ClusterONE achieved comparable accuracies when matched against known `gold standard' complexes and MCL achieved the closest performance to ClusterONE with the exception that MCL produced only non-overlapping clusters - a distinct advantange of ClusterONE.

\subsubsection*{Some other methods based on graph clustering}
Apart from these discussed methods, three other methods worth mentioning here are LCMA (2005)\cite{Li2005}, 
PCP (2007)\cite{Chua2007} and HACO (2009)\cite{Wang2009}.
The LCMA algorithm first locates cliques within local neighborhoods using vertex degrees and then merges them based
on overlaps to produce complexes. Protein Complex Prediction (PCP)
uses FS Weight scoring to remove unreliable interactions and add indirect interactions, and then merges cliques
to produce the final list of complexes. On the other hand, HACO uses hierarchical agglomerative clustering
to produce the intial set of (non-overlapping) clusters. Proteins are then assigned to multiple clusters based on their
interactions to the clusters to produce the final list of overlapping clusters.

A few other recently proposed (2010 - 2011) methods include those by Zhang et al.\cite{Zhang2011}, Ma et al.\cite{Ma2011}, 
Wang et al.\cite{Wang2011} and Chin et al.\cite{Chin2010}.
These use the property of ``bridgeness" of cross-edges among clusters along with the 
internal connectivities to detect complexes.

\subsection{Methods incorporating core-attachment structure}
Gavin and colleagues (2006)\cite{Gavin2006} performed large-scale analysis of yeast complexes and found
that the proteins with complexes were divided into two distinct groups, ``cores" and ``attachments".
The cores formed central functional units of complexes, while the attachment proteins aided these cores
in performing their functions. Several computational methods were proposed to reconstruct complexes from PPI newtorks
by capitalizing on this structural organization.

\begin{figure}[th]
\centerline{\psfig{file=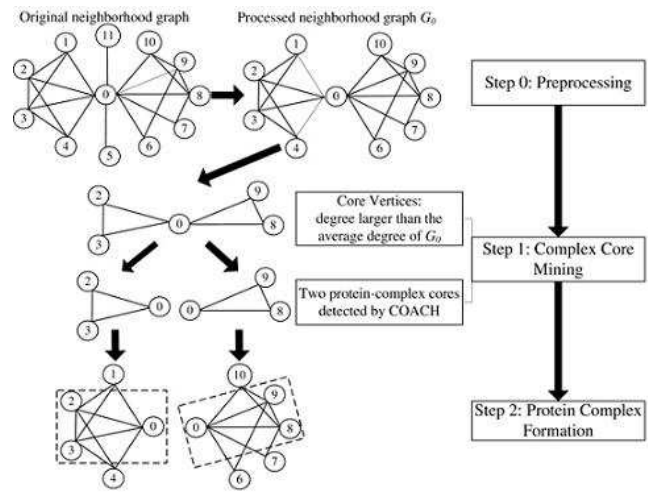,width=8.5cm}}
\vspace*{8pt}
\caption{The identification of core and attachment proteins in COACH:
The cores are first identified based on vertex degrees in the neighborhood graphs. Attachment proteins are
then appended to these cores to build the final complexes.}
\label{figure:COACH_XLi2010}
\end{figure}

Wu Min et al. (2009)\cite{WuMin2009} proposed the COACH method which reconstructs complexes in two stages - it identifies
dense core regions, and subsequently includes proteins as attachments to these cores. Figure~\ref{figure:COACH_XLi2010} 
summarizes how COACH identifies core and attachment proteins to build complexes.

Leung et al. (2009)\cite{Leung2009} proposed the CORE method to identify protein cores within the PPI network.
They defined the probability of two proteins $p_1$ and $p_2$ (of degrees $d_1$ and $d_2$, respectively)
to belong to the same core using two main factors: 
whether the two proteins interact or not and the number of common neighbors $m$ between them. 
The probability that $p_1$ and $p_2$ have $\geq i$ interactions and $\geq m$ common neighbors is calculated
under the null hypothesis that $d_1$ edges connecting $p_1$ and $d_2$ edges connecting $p_2$ are randomly
assigned in the PPI network according to a uniform distribution. This probability is used to arrive at a 
$p$-value for whether $p_1$ and $p_2$ belong to the same core.
Subsequently, CORE merges sets of core proteins of sizes two, three, etc. until further increase in size
is not possible, to produce the final set of cores.
CORE then scores and ranks the predicted cores based on the number of internal and external interactions 
in them. The attachments are added to these cores in a manner similar to COACH to produce the final set of complexes.

Srihari et al. proposed MCL-CA (2009)\cite{Srihari2009} and the improved (weighted) version MCL-CAw (2010)\cite{Srihari2010} 
which identify complexes by refining clusters produced by the MCL algorithm\cite{Enright2002,Enright2004} by 
incorporating core-attachment structure.
Essentially, MCL-CAw categorizes proteins within MCL clusters into ``core" and ``attachment" proteins based on their connectivities,
and then selects only these categorized proteins into complexes while discarding the remaining ``noisy" proteins. This
enables MCL-CAw to ``trim" the raw MCL clusters. Further, unlike CORE and COACH, the refinement in MCL-CAw
capitalizes on reliability scores assigned to the interactions.
Consequently, MCL-CAw reconstructs significantly higher number of `gold standard' complexes 
($\sim$30\% higher) and with better accuracies compared to plain MCL.

Chin et al. (2010)\cite{Chin2010} proposed the HUNTER algorithm which begins by generating a module seed $MS(v)$ for
each node $v$ in the PPI network. $MS(v)$ is then pruned by removing vertices having low weight edges to other
members of $MS(v)$. Then the maximal $q$-connected subnetwork of $MS(v)$ is selected as the initial core $MQC(v)$. 
This core is then expanded into a module by adding new vertices that share many neighbors with $MQC(v)$. If two
modules overlap beyond a certain threshold, these modules are merged. The resultant collection of modules form
the final set of predicted complexes.

\subsection{Methods incorporating functional information}
Proteins within complexes are generally enriched with same or similar functions\cite{Zhang2008,Gavin2006}.
If the functional information for proteins from an organism are available, then this information can be combined with topological
information from PPI networks for the reconstruction of complexes from the organism.
One possible way to incorporate functional information is to score the interactions based on the
functional similarity between the interacting pairs of proteins. 
Alternately, functional annotations (for example, from Gene Ontology\cite{Ashburner2000})
can be used to aid decisions where including or excluding a protein 
into complexes purely based on topological information might be difficult.

\subsubsection*{Restricted Neighborhood Search Clustering (RNSC)}
King et al. (2004)\cite{King2004} proposed the RNSC algorithm that combines topological and Gene Ontology
information to detect complexes. The algorithm operates in two steps - it begins by clustering the PPI network
and then filters the clusters based on cluster properties and functional homogeneity. 

The network $G=(V,E)$ is first randomly partitioned into multiple subnetworks, which is essentially
a partitioning of the node set $V$. The algorithm then iteratively moves nodes from one cluster to another
in a randomized fashion till an integer-valued cost function is optimized. A common problem among such
clustering algorithms is the tendency to settle in poor local minima. To avoid this, the RNSC algorithm
adopts diversification moves, which shuffle the clustering by occasionally dispersing the contents of
a cluster at random. Once the clustering process is completed, clusters of small sizes or densities (the lower bound
on cluster sizes and densities are experimentally determined) are discarded. Finally, a $p$-value is calculated
using functional annotations (from GO) for each cluster that measures the functional homogeneity of the clusters.
All clusters above a certain $p$-value are discarded to produce the final list of predicted complexes.
Based on experiments, King et al. recommend cluster density cut-off of 0.70 and $p$-value cut-off of $10^{-3}$.

\subsubsection*{Dense neighborhood Extraction using Connectivity and conFidence Features (DECAFF)}
Li et al. (2007)\cite{Li2007} proposed the DECAFF algorithm which essentially is an extention of the LCMA algorithm\cite{Li2005}
proposed earlier by the same group. DECAFF identifies dense subgraphs in a neighborhood graph using a hub-removal algorithm.
Local cliques are identified in these dense subgraphs and merged based on overlaps
to produce clusters. 
Each cluster is assigned a functional reliability score, which is the average of the reliabilities of the edges
within the cluster. All clusters with low reliabilities are discarded to produce the final set of predicted complexes.

The PCP algorithm\cite{Chua2007} described earlier can also be categorized into this set of methods because PCP
uses a weighting scheme based on functional similarity (though the similarity is inferred from topology)
to assign reliability scores to interactions, and then uses a clique merging strategy to detect complexes.

\subsection{Methods incorporating evolutionary information}
The increasing availability of PPI data from multiple species like yeast, fly, worm and some mammals has made
it feasible to use insights from cross-species analysis for detection of (conserved) complexes. The assumption is
that proteins belonging to real complexes should generally be conserved through the evolution process to act 
as an integrated functional unit\cite{Zhang2008}.

Sharan et al. proposed methods (2005-2007)\cite{Sharan2004,Sharan2006} for detection of conserved complexes across species
based on the evolution of PPI networks. In these methods, an orthology network (network alignment graph) is constructed from the
PPI networks of different species, which essentially represents the orthologous proteins and their
conserved interactions across the species. For a protein pair $\{u_1, v_1\}$ in network $G_1$ of species $S_1$ and $(u_2, v_2)$
in $G_2$ of species $S_2$, the orthology network $G_{12}$ contains the pair $\{u, v\}$ if $u_1$ is orthologous to $u_2$,
and $v_1$ is orthologous to $v_2$. The edge $(u,v)$ is weighted by the sequence similarities between the pairs $\{u_1, v_1\}$,
and $\{u_2,v_2\}$. Any subgraph in this orthology network $G_{12}$ is therefore a conserved subnetwork of $G_1$ and $G_2$.
Such candidate subgraphs are then evaluated for parts of conserved complexes. 
Based on this idea, a tool \emph{QNet}\cite{Sharan2007} was developed which returns conserved complexes from 
different species when queried using known complexes from yeast.

\subsection{Methods based on co-operative and exclusive interactions}

The overlapping binding interfaces in a protein may prevent multiple interactions involving these interfaces
from occurring simultaneously\cite{Kim2006}. In other words, the set of interactions in which a protein participates may
be either co-operative or mutually exclusive. The information about the co-occurrence or 
exclusiveness of interactions can therefore be useful for predicting complexes with higher accuracy.
This information can be gathered from the interacting domains of protein pairs
or the three-dimensional structures of the interacting surfaces.

Ozawa et al. (2010)\cite{Ozawa2010} proposed a refinement method over MCODE and MCL to filter predicted 
complexes based on exclusive and co-operative interactions. They used domain-domain interactions 
to identify conflicting pairs of protein interactions in order
to include or exclude proteins within candidate complexes. 
Based on their results, the accuracies of predicted complexes from MCODE and MCL improved by two-fold.

On the other hand, Jung et al. (2010)\cite{Jung2010} used structural interface data to construct a simultaneous
PPI network (SPIN) containing only co-operative interactions and excluding competition from mutually exclusive
interactions. MCODE and LCMA algorithms tested on this SPIN displayed
a sizeable improvement in correctly predicted complexes.

Even though incorporating information about co-operative and exclusive interactions shows promising
improvement in complex detection algorithms, there are still several practical problems related to this approach.
Gathering more data on conflicting interactions, especially based on three-dimensional structures of interfaces, 
needs to be addressed before this approach can be more easily adopted.

\subsection{Incorporating other possible kinds of information}
In a recent foresightful survey by Przytycka et al.\cite{Przytycka2010}, the application of network \emph{dynamics}
(temporal information) into current computational analysis is discussed at good lengths, especially with respect
to detection of complexes and pathways from protein interaction networks. The authors suggest that
if sufficient information about the `timing activities' of proteins can be obtained, the dynamical nature of the underlying 
organizational principles in interaction networks can be better understood.
This shift from static to dynamic network analysis is vital to understanding several
cellular processes, some of which may have been wrongly understood due to ignoring dynamic information.

\section{Comparative assessment of existing methods}
Considering the wide variety of proposed methods for complex detection, one can gauge 
the seriousness in the research effort towards computational identification and categorization of complexes.
Several surveys and experiments\cite{Brohee2006,Vlasblom2009,Xli2010}
have focused on the comparative analyses of these proposed methods for complex detection.
Each new work on complex detection also comes with detailed comparative analyses of the new
method with some earlier methods. However, due to the differences in PPI and benchmark datasets, evaluation
criteria, thresholds and parameters used, and the subset of methods considered for these comparative assessments,
different works arrive at different results on the relative performance of methods. But, typically the following broadly accepted
criteria are used across the works.

If a reasonably large `gold standard' set of complexes is available (as in the case of yeast),
the performance of a method can be gauged on how accurately its predicted complexes reconstruct or recover the `gold standard'. Two commonly adopted
measures for this are precision and recall\cite{Liu2009}. 
Precision measures how many among the predicted complexes match some `gold standard' complex, in turn measuring
the proportion of realible predictions (accuracy) from the method. 
Recall measures how many of the `gold standard' complexes are reconstructed by the method, in turn measuring the coverage or sensitivity of the method.
Some methods tend to produce too many (sometimes arbitrary) predictions resulting in high recall but very low precision, and therefore too many false positives to
consider the method even reasonably reliable. To handle this,
a combination of precision and recall, usually through a harmonic mean called F-measure, is used to evaluate how ``balanced" is the method.

On the other hand, if a `gold standard' set is not available (as in the case mammals, currently), ``self-evaluatory" measures like cluster cohesiveness and
separability is used\cite{Bader2003,King2004}. The cohesiveness of a predicted complex (cluster) usually measures topological
characteristics of the cluster, for example, its interaction density or size, while separability measures how separated is the cluster from others\cite{Bader2003}.
A combination of cohesiveness and separability reveals how modular is the clustering and therefore how likely the individual clusters represent real complexes.

Another typically independent way to evaluate the predictions is to measure the functional or co-localization coherence of the clusters subjected, however, to availability of appropriate annotation data\cite{Liu2009,King2004}.
This captures how functionally coherent are the proteins within a predicted complex and whether they are co-localized within the cell.
The usual annotations required for these calculations are functions and sub-cellular localizations of the proteins.
This evaluation is particularly useful for alternative validation of the predictions.

Now, we present a summary of some representative surveys and comparative assessments and their conclusions.
One of the first comprehensive assessments was performed by Brohee and van Helden (2006)\cite{Brohee2006}. 
They performed a detailed empirical comparison between MCODE\cite{Bader2003},
MCL\cite{Dongen2000}, RNSC\cite{King2004} and Super-paramagnetic Clustering (SPC)\cite{Blatt1996}.
These algorithms were tested on PPI datasets from high-throughput experiments, and the resultant complexes
were evaluated against benchmark complexes from MIPS\cite{Mewes2006}. Additionally, the PPI datasets
were introduced with artificial noise (random edge addition and deletion) to test the robustness of these
algorithms.
They concluded that MCL and RNSC outperformed MCODE and SPC in terms of precision (the proportion of correctly
predicted complexes) and recall (the proportion of correctly derived benchmarks). RNSC was robust to variation
in its input parameter settings, while the performance of the other three varied widely for parameter changes.
MCL was remarkably robust even upon introducing 80\%-100\% random noise. Overall, the experiments confirmed
the general superiority of MCL over the other three algorithms.

Vlasblom et al. (2009)\cite{Vlasblom2009} compared MCL with another clustering algorithm,
Affinity Propagation (AP)\cite{Frey2007} on unweighted as well as weighted PPI networks.
The initial unweighted network was built from a set of 408 hand-curated complexes from Wodak lab\cite{Pu2009}
followed by random addition and removal of edges to mimic real PPI networks. The weighted network
was obtained from the Collins et al.'s work\cite{Collins2007}, generated from Gavin
and Krogan datasets\cite{Gavin2006,Krogan2006}. 
They concluded that MCL performed considerably better than AP
in terms of accuracy and separation of predicted clusters, and robustness to random noise.
In particular, MCL was able to achieve about 90\% accuracy and 80\% separation compared to only
70\% accuracy and 50\% separation of AP on unweighted PPI networks with introduced random noise.
MCL was able to discover benchmark complexes even at high (40\%) noise levels.

\begin{figure}
\centerline{\psfig{file=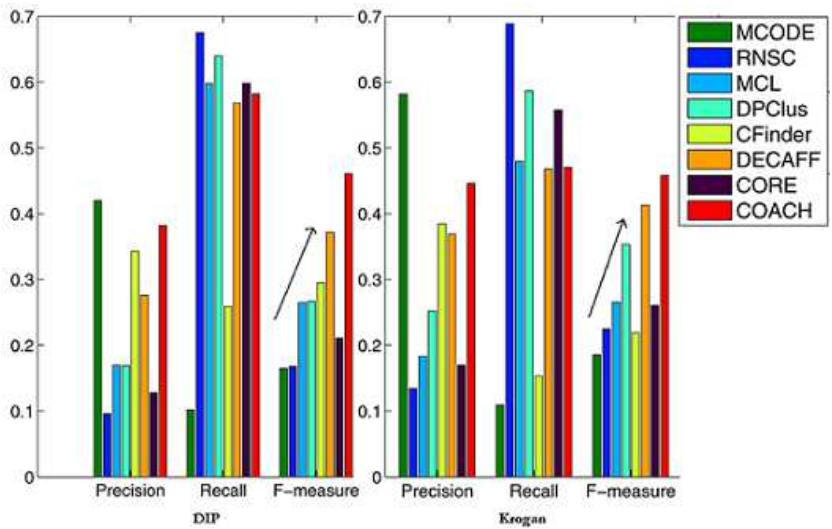,width=11cm}}
\vspace*{8pt}
\caption{Comparative performance of complex detection methods in terms of precision, 
recall and F-measure on DIP and Krogan datasets. The methods are arranged in chronological order, and it is interesting
to note that over the years, the F-measures have improved.}
\label{figure:Comparison_methods_DIP_Krogan_XLi2010}
\end{figure}

More recently (2010), Li et al.\cite{Xli2010} performed a detailed comparative evaluation of several algorithms:
MCODE\cite{Bader2003},
MCL\cite{Dongen2000}, CORE\cite{Leung2009}, COACH\cite{WuMin2009}, RNSC\cite{King2004} and DECAFF\cite{Li2007}.
These algorithms were tested on PPI datasets from DIP\cite{Xenarios2002}
and Krogan et al.\cite{Krogan2006}. The DIP network consisted of 17203 interactions among 4930 proteins, while
the Krogan dataset consisted of 14077 interactions among 3581 proteins. They used a total of 428 benchmark
complexes from MIPS\cite{Mewes2006}, Aloy et al.\cite{Aloy2004} and SGD\cite{Cherry98}. A cluster $P$ from
a method was considered a correct match to a benchmark complex $B$ using the Bader score\cite{Bader2003} $|V_P \cap V_B|^2 / (|V_P|.|V_B|) \geq 0.20$,
where $V_P$ denotes the number of proteins in $P$, and $V_B$ denotes the number of proteins in $B$.
Based on this criteria, the precision, recall and F-measure values were calculated.
The comparisons of these algorithms is shown in Figure~\ref{figure:Comparison_methods_DIP_Krogan_XLi2010}
(adapted from\cite{Xli2010}). The methods are arranged in chronological order, and it is interesting
to note that over the years, the F-measures have improved.
Li et al. concluded that MCL, RNSC, CORE, COACH and DECAFF attained the best recall
values. MCODE was able to achieve the highest precision, but it produced very few clusters resulting in very low recall.

%%%%%%%%%%%

\subsection{Our assessment of some complex detection methods}

\subsubsection{Preparation of experimental data}
In our assessment, we experimentally evaluated some key complex detection methods on both unscored (raw) and scored PPI networks.
To build our unscored network, we combined the physical interaction from two TAP-MS experiments, Gavin et al. (2006)\cite{Gavin2006}
and Krogan et al. (2006)\cite{Krogan2006}, which we call the Gavin+Krogan (G+K) network. We then gathered the scored version
of this network, the Consolidated network from Collins et al. (2007)\cite{Collins2007}. This network comprises of interactions
from Gavin et al. and Krogan et al. scored using the Purification Enrichment scheme\cite{Collins2007}. Some of the properties of these
networks are shown in Table 1.

%%%%%%%
%%%PASTE TABLE HERE
%%%%%%%%

\begin{table*}[ht]
  \begin{center}
  {\small
  \begin{tabular}{ c || c | c | c }
  \hline
	PPI Network	 			& {\# Proteins} & {\# Interactions} &   {Avg node degree} \\[0.8ex]
	\hline	
	Gavin	      				& 1430           & 7592 		& 10.62\\[0.8ex]
	Krogan `Core'	 			& 2708 		 & 7123			& 5.26 \\[1.0ex]
	\hline
	Gavin+Krogan  				& 2964		 & 13507		& 9.12 \\[0.8ex]		
	Consolidated				& 1622		 & 9704			& 11.96\\[0.8ex]		
  \hline
  \end{tabular}
  }
  \end{center}
  \caption{Properties of the PPI networks used for the evaluation of methods}
  \label{PPI_Network_Properties}
  \end{table*}

The benchmark (reference or `gold standard') set of complexes was built from three independent sources:
408 complexes of the Wodak lab CYC2008 catalogue\cite{Pu2009},
313 complexes of MIPS\cite{Mewes2006}, and
101 complexes curated by Aloy et al.\cite{Aloy2004}.
The properties of these reference sets are shown in Table 2.
We considered each of these reference sets independently for the evaluation of the methods.
We did not merge them into one comprehensive list of complexes because 
the individual complex compositions are different across the three sources 
and some complexes may also get double-counted (because of different names used for the same complex).

%%%%%%%
%%%PASTE TABLE HERE
%%%%%%%%

   \begin{table*}[ht]
    \begin{center}
    {\small
    \begin{tabular}{ c || c || c || c | c | c | c || c }
    \hline
    			&			&		& \multicolumn{4}{c ||}{\em \# Complexes of size} & 		\\
    {Benchmark}		& {\#Complexes} 	& {\# Proteins}	& $<$ 3  & 3-10		& 11-25		& $>$ 25 & {Avg density}\\
    \hline
    Wodak			& 408			& 1627		& 172 	 & 204		& 27		& 5	 &  0.639\\
    	
    MIPS			& 313			& 1225		& 106    & 138		& 42		& 27	 &  0.412\\
    
    Aloy			& 101			& 630		& 23     & 58		& 19		& 1      &  0.747\\
     \hline
     
  \end{tabular}
  }
  \end{center}
  \caption{Properties of hand-curated (\emph{bona fide}) 
  yeast complexes from Wodak lab, MIPS and Aloy}
  \label{Size_distribution_benchmarks}
  \end{table*}

\subsubsection{Metrics for evaluating the predicted complexes}
Let $\mathcal{B} = \{B_1, B_2,...,B_m \}$ and $\mathcal{C} = \{C_1, C_2, ...,C_n \}$ be the sets of 
benchmark and predicted complexes, respectively.
We use the Jaccard coefficient $J$ to quantify the overlap between a benchmark complex $B_i$ and a predicted complex $C_j$:

\begin{equation}
J(B_i,C_j) = \frac{|B_i \cap C_j|}{|B_i \cup C_j|}.
\end{equation}

\noindent We consider $B_i$ to be covered by $C_j$, if $J(B_i,C_j) \geq$ {\em overlap threshold} $t$. In our experiments,
we set the threshold $t = 0.5$, which requires $|B_i \cap C_j| \geq \frac{|B_i| + |C_j|}{3}$. For example,
if $|B_i| = |C_j| = 8$, the overlap between $B_i$ and $C_j$ should be at least 6.

We use previously reported\cite{Liu2009} definitions of {\em recall} (coverage) and {\em precision} (sensitivity)
of the set of predicted complexes:

\begin{equation}
\label{eq_recall}
	Recall~Rc = \frac{|\{B_i|B_i \in \mathcal{B} \wedge \exists C_j \in \mathcal{C}; J(B_i,C_j) \geq t\}|}{|\mathcal{B}|}
\end{equation}

\noindent Here, $|\{B_i|B_i \in \mathcal{B} \wedge \exists C_j \in \mathcal{C}; J(B_i,C_j) \geq t\}|$ gives the number of {\em derived benchmarks}.

\begin{equation}
\label{eq_precision}
	Precision~Pr = \frac{|\{ C_j| C_j \in \mathcal{C} \wedge \exists B_i \in \mathcal{B}; J(B_i,C_j) \geq t \} |}{|\mathcal{C}|}
\end{equation}

\noindent Here, $|\{C_j| C_j \in \mathcal{C} \wedge \exists B_i \in \mathcal{B}; J(B_i,C_j) \geq t\}|$ gives the number of {\em matched predictions}.

We calculated the {\em F-measure} as the harmonic mean of precision and recall,
\begin{equation}
F = \frac{2*Pr*Rc}{Pr + Rc}
\end{equation}

%%%%%%%%

\subsubsection{Experimental evaluation of methods}
We considered the following methods for our evaluation: 
\begin{itemize}
\item On the unscored network, MCL (2002, 2004)\cite{Dongen2000,Enright2004}, MCL-CA (2009)\cite{Srihari2009}, MCL-CAw (2010)\cite{Srihari2010}, CORE (2009)\cite{Leung2009},
COACH (2009)\cite{WuMin2009}, CMC (2009)\cite{Liu2009} and HACO (2009)\cite{Wang2009};
\item On the scored network, MCL (2002, 2004)\cite{Dongen2000,Enright2004}, MCLO (2007)\cite{Pu2007}, MCL-CAw (2010)\cite{Srihari2009,Srihari2010}, CMC (2009)\cite{Liu2009} and HACO (2009)\cite{Wang2009}.
\end{itemize}
We do not show comparisons for older methods like MCODE (2003)\cite{Bader2003} and RNSC (2004)\cite{King2004} because these have been
evaluated extensively in several earlier surveys\cite{Brohee2006,Xli2010}, instead we included MCL as a benchmark in all our comparisons since MCL
has been repeatedly shown to perform better than these older methods\cite{Brohee2006,Vlasblom2009,Xli2010}.
Further, not all methods are devised to make use of interaction confidence scores, and therefore we selected only the ones capable of doing so
for the evaluations on the scored network.

Table 3 shows the precision and recall values for methods evaluated on the unscored
Gavin+Krogan network across the three benchmark sets. The table shows that CORE, HACO and MCL-CAw performed significantly better
in terms of recall compared to the rest of the methods. In particular, MCL-CAw performed considerably better than plain MCL indicating
that incorporating core-attachment structure into raw MCL clusters helped to improve the accuracies of the predicted complexes. This indicated
that incorporating some kind of biological knowledge helped to identify complexes more accurately.

Next, Table 4 shows these values for the methods evaluated on the scored Consolidated
network. This table shows that all methods were able to reconstruct significantly higher number of complexes upon
scoring as compared to on the unscored network. This clearly indicated that noise in raw datasets (negatively) impacted
the performance of methods, and reliability scoring aided in alleviating this impact and thereby improving the performance of methods.
This demonstrated the effectiveness of current reliability scoring schemes in cleaning raw interaction datasets 
for focused studies such as complex detection.

%%%%%%%%%%
%%TABLES PASTE HERE
%%%%%%%%%%%%

\begin{table*}[htp]
\begin{center}
{\scriptsize
\begin{tabular}{ c || l || c | c | c | c | c | c | c }
\multicolumn{9}{c}{{\em The unscored Gavin+Krogan network}}\\[0.8ex]
\multicolumn{9}{c}{{\#Proteins 2964; \#Interactions 13507}}\\[1.2ex]
\hline
& 	  	 & \multicolumn{7}{c}{\em Method}\\[0.8ex]
		&  				& {MCL}  	& {MCL-CA}	& {MCL-CAw}   		& {COACH}		&{CORE}		&{CMC}		& {HACO}\\[0.8ex]
\cline{3-9} 
		&  {\#Predicted}  		& 242      	& 219		& 130	      		& 447			& 386		& 113		& 278	\\[0.8ex]
\hline

		& {\#Matched}  			& 55  		& 49		& 69	 		& 62			& 83		& 60		& 78	\\[1.0ex]

\bf{Wodak}	& {Precision} 			& 0.226    	& 0.224		& 0.531 		& 0.139			& 0.215		& 0.531		& 0.281	\\[0.8ex]   

\bf{(\#182)}	& {\#Derived}    		& 62  	 	& 49		& 75  			& 49			& 83		& 60		& 85	\\[0.8ex]

		& {Recall}       		& 0.338    	& 0.269		& 0.412 		& 0.269			& 0.456		& 0.330		& 0.467	\\[1.2ex]

\hline

		& {\#Matched}  			& 35   		& 42		& 42  			& 45			& 59		& 41		& 45	\\[0.8ex]

\bf{MIPS}	& {Precision} 			& 0.143    	& 0.192		& 0.323 		& 0.101			& 0.153		& 0.363		& 0.162	\\[0.8ex]

\bf{(\#177)}	& {\#Derived}			& 40   		& 42		& 53  			& 38			& 59		& 41		& 57	\\[0.8ex]

		& {Recall}       		& 0.226    	& 0.237		& 0.300 		& 0.215			& 0.333		& 0.232		& 0.322	\\[1.2ex]

\hline

		& {\#Matched}  			& 43  		& 41		& 47  			& 54			& 59		& 43		& 59	\\[0.8ex]

\bf{Aloy}	& {Precision} 			& 0.179    	& 0.187		& 0.362 		& 0.121			& 0.153		& 0.381		& 0.212	\\[0.8ex]

\bf{(\#76)}	& {\#Derived}    		& 42   		& 41		& 52  			& 37			& 59		& 43		& 59	\\[0.8ex]

		& {Recall}       		& 0.556    	& 0.539		& 0.684 		& 0.487			& 0.776		& 0.566		& 0.776	\\[1.2ex]

\hline

\end{tabular}
}
\end{center}
\caption{Comparisons between different methods on the unscored Gavin+Krogan network. CORE showed the best recall
followed by HACO and MCL-CAw.}
\label{Comparison_algorithms_Gavin+Krogan}
\end{table*}

\begin{table*}[ht]
\begin{center}
{\scriptsize
\begin{tabular}{ c || l || c | c | c | c | c }
\multicolumn{7}{c}{{\em The Consolidated$_{3.19}$ network}}\\[0.8ex]
\multicolumn{7}{c}{{\#Proteins 1622; \#Interactions 9704}}\\[1.2ex]
\hline
& 	  	 & \multicolumn{5}{c}{\em Method}\\[0.8ex]
		&  				& {MCL}  		& {MCLO}			& {MCL-CAw}   			&{CMC}			& {HACO}\\[0.8ex]
\cline{3-7} 
		&  {\#Predicted}  		& 116      		& 119				& 130 	      			& 77			& 101	\\[0.8ex]
\hline

		& {\#Matched}  			& 70  			& 80				& 83	 			& 67			& 57	\\[1.0ex]

\bf{Wodak}	& {Precision} 			& 0.603    		& 0.672				& 0.638 			& 0.870			& 0.564	\\[0.8ex]   

\bf{(\#145)}	& {\#Derived}    		& 79  	 		& 80				& 90  				& 67			& 64	\\[0.8ex]

		& {Recall}       		& 0.545    		& 0.552				& 0.621 			& 0.462			& 0.441	\\[1.2ex]

\hline

		& {\#Matched}  			& 48   			& 65				& 53  				& 56			& 40	\\[0.8ex]

\bf{MIPS}	& {Precision} 			& 0.414    		& 0.546				& 0.408 			& 0.727			& 0.396	\\[0.8ex]

\bf{(\#157)}	& {\#Derived}			& 63   			& 65				& 67  				& 56			& 57	\\[0.8ex]

		& {Recall}       		& 0.401    		& 0.414				& 0.427 			& 0.357			& 0.363	\\[1.2ex]

\hline

		& {\#Matched}  			& 54  			& 56				& 57  				& 45			& 44	\\[0.8ex]

\bf{Aloy}	& {Precision} 			& 0.466    		& 0.471				& 0.438 			& 0.584			& 0.436	\\[0.8ex]

\bf{(\#76)}	& {\#Derived}    		& 55   			& 56				& 55  				& 45			& 45	\\[0.8ex]

		& {Recall}       		& 0.724    		& 0.737				& 0.724 			& 0.592			& 0.592	\\[1.2ex]

\hline

\end{tabular}
}
\end{center}
\caption{Comparisons between the different methods on the Consolidated$_{3.19}$ network. MCL-CAw showed the best recall followed by CMC.}
\label{Comparison_algorithms_Consolidated$_{3.19}$}
\end{table*}

\subsubsection{Plugging experimental results into our taxonomy}
\begin{figure}[th]
\centerline{\psfig{file=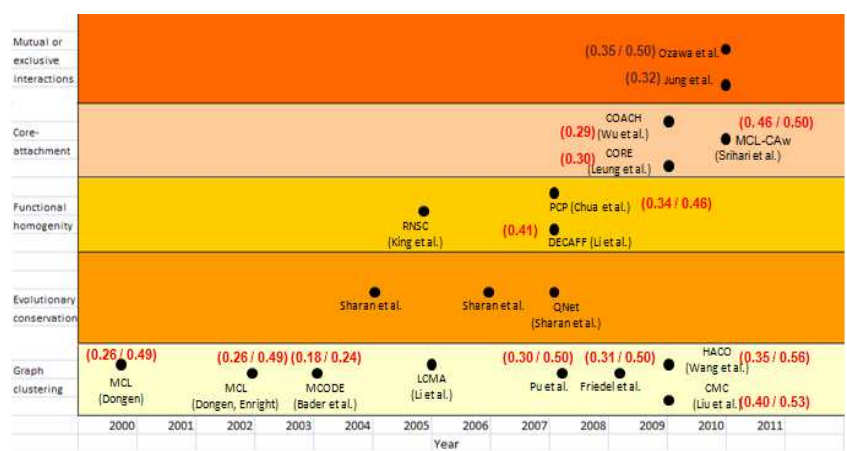,width=11.5cm,height=6cm}}
\vspace*{8pt}
\caption{Plugging-in F-values (before-after scoring) of existing methods into our Bin-and-Stack classification.
Incorporating biological information and affinity scoring significantly boosts performance.}
\label{figure:Taxonomy_chronology_F1}
\end{figure}

We next ``plugged-in" these evaluation results as well as results obtained from some earlier surveys\cite{Vlasblom2009,Xli2010} into our
``bin-and-stack" classification, as shown in Figure~\ref{figure:Taxonomy_chronology_F1}. 
For each method, we show the F-values \emph{before / after} scoring, that is, on the unscored G+K network and the scored
Consolidated network.  
Such a representation in our classification revealed two interesting insights,
\begin{enumerate}
\item incorporating biological information in addition to PPI topology improved performance of the methods 
(the F measures have increased in the vertical layers compared to the lowest layer);
\item reliability scoring significantly improved performance of the methods, as shown by the before-after values.
\end{enumerate}
This representation also shows how complex detection methods have evolved over the years to improve performance, and therefore
our taxonomy can be useful to guide future directions for further improvement.

\section{Open challenges in complex detection from PPI networks}
The review and evaluation of computational methods in the above sections reveal several critical challenges facing accurate 
identification of complexes from high-throughput interaction datasets. We saw that most methods are
considerably impacted by noise in raw datasets.
Further, most methods are able to reconstruct only a fraction of the known complexes (achieve at the most $65\%$ recall) 
even upon scoring. This points towards some inherent limitations within the methods itself. 
On this basis, we broadly classify the challenges facing current methods into two categories,
(i) challenges originating from biological datasets;
(ii) challenges originating from existing computational techniques.

\subsection{Challenges from interaction datasets}
Even though over the last few years, several independent high-throughput experiments
have helped to catalogue enormous amount of interactions from yeast, 
they show surprising lack of correlation with each other, and lack of coverage - 
bias towards high abundance proteins and against proteins
from certain cellular compartments (like cell wall and plasma membrane)\cite{Bader2002,vonMering2002,Cusick2008,Mackay2008}.
Also, each dataset still contains a substantial number of false positives (noise) that can compromise the
utility of these datasets for more focused studies like complex detection, as seen from our evaluation results.
In order to reduce the impact of such discrepancies, a number of data integration and reliability scoring schemes 
have been devised\cite{Collins2007,Hart2007,Chua2008,Liu2008,Friedel2008,Kuchaiev2009,Voevodski2009,Jain2010,Suthram2006,SrihariThesis2012}.

To overcome these challenges to some extent, in our evaluation, we combined multiple datasets 
(from Gavin et al.\cite{Gavin2006} and Krogan et al.\cite{Krogan2006}) to account for the lack of interaction coverage, 
and also adopted scoring prior to predicting complexes.
In spite of these precautionary steps, we notice that most methods are able to reconstruct only a fraction of the known complexes, and
we still have a long way to go towards identification of meaningful novel complexes through computational means.

\subsection{Challenges from existing complex detection methods}
As noted earlier, even though there have been numerous methods developed for complex detection, most of them suffer from low recall
(at most 65\% recall even on the scored network; Table 4).
Even a ``union" of these methods achieves at most $70$-$75\%$ recall on average across a 
variety of PPI datasets\cite{Srihari2012}.
One of the crucial reasons for this limitation is that every method, in one way or another, 
relies on the key assumption that complexes are embedded among ``dense" regions of the network. 
However, recent experiments have indicated that relying too much on this assumption in the wake of
insufficient credible interaction data causes these methods to miss many complexes that are of low densities, 
that is, ``sparse" in the network\cite{Srihari2012}. Therefore the need is to find alternative ways to model complexes than mere
dense subnetworks, and also to compensate for the sparsity of topological information by augmenting 
other kinds of biological information.

\subsubsection{Detection of sparse complexes}
In the attempt to detect sparse complexes, the recent work by Srihari et al. (2012)\cite{Srihari2012} is insightful and can guide
futher directions towards tackling this challenge.
Srihari et al. (2012)\cite{Srihari2012} noticed that most existing computational methods based on PPI networks 
rely overly on the assumption that complexes
are embedded among ``dense" regions of the network, and therefore miss most of the ``sparse" complexes that have very low
interaction densities or lie disconnected in the network. These complexes are missed by current methods due to the lack of crucial
topological information (missing interactions and/or proteins) - for example, even in the well-studied organism yeast, only $\sim 70\%$ of the
interactome has been validated and catalogued\cite{Cusick2008}.
In order to overcome these ``topological gaps", the authors proposed careful augmenting of functional
interactions to PPI networks. Functional interactions represent logical or conceptual associations among proteins, and therefore
``encode" a variety of biological similarities or affinities among proteins beyond just physical interactivity, 
thereby compensating for the lack of topological information in PPI networks.
In order to do this augmentation systematically they proposed the SPARC algorithm. 

SPARC selects low quality clusters predicted from the physical (PPI) network using existing methods, and then selectively
enhances these clusters to reconstruct (sparse) complexes. The key idea is that many of these low quality clusters are in fact fractions (or ``pieces") of
sparse complexes embedded in the PPI network, but due to missing interactions they lie ``scattered" in the network and 
do not represent whole complexes in their current forms.
If these clusters can be carefully enhanced by augmenting functional interactions, then several of the sparse complexes can be reconstructed
accurately. This enhancement involves increasing their interaction densities and ``pulling-in" together their disconnected components.
However, during the selection of the initial set of low quality clusters, many may just represent noisy predictions (false positives).
In order to determine the clusters that are more likely to represent complexes, SPARC makes use of a novel Component-Edge (CE) score.
The CE-score is a topological measure combining connectivity and relative density of the clusters, and is shown to more accurately correlate with
the topological characteristics of real complexes compared to traditionally accepted measures like edge density.
The CE-score is calculated for every low quality cluster predicted from the PPI network. SPARC then augments functional interactions to
the clusters and checks if their CE-scores improve beyond a certain threshold. If a cluster shows the required improvement, it
is output as a potential candidate representing a sparse complex.
Srihari et al. showed through extensive experiments on clusters produced from methods like 
MCL\cite{Dongen2000,Enright2004}, MCL-CAw\cite{Srihari2010}, CMC\cite{Liu2009} and HACO\cite{Wang2009} that SPARC was 
capable of improving the overall recall of these methods by upto $47\%$ on average across a variety of networks. 
Specifically, SPARC helped to reconstruct $25\%$ more complexes among the ones that were sparse.

\subsubsection{Detection of small and temporal complexes}
Small complexes (complexes of two or three proteins) also pose severe challenges in identification, particularly if PPI network topology is the 
only available information. In fact most complex detection methods based solely on the PPI network only attempt to identify complexes
with at least 4 or more proteins in the network\cite{Liu2009,Leung2009,Srihari2010}. The attempt to predict small complexes
(pairs or triplets) from the network based only on connectivity typically produces a significant fraction of false positives.
Further, the smaller the size of a complex, the more prone it is to sparsity - 
missing even a few interactions can result in very low densities or render the complexes disconnected.
Due to these challenges, additional information apart from PPI network topology is required for detection of such complexes.

In a recent work (2012), Srihari et al.\cite{Srihari2012poster} incorporated temporal information to identify small complexes. 
The authors focused on identifying small complexes that are assembled during the yeast cell cycle in a temporal
manner. They noticed that several high-degree proteins such as kinases interacted with different subsets of proteins during different
phases of the yeast cell cycle to assemble multiple phase-specific complexes.
However, due to the lack of temporal information to disambiguate (segregate) the different phase-specific complexes, existing
topology-based methods produced large clusters of complexes fused together. Srihari et al.
decomposed such large clusters by incorporating temporal information on the yeast cell cycle in which each protein showed peak expression,
and thereby segregated the individual constituent complexes. Many of the segregated complexes were small and represented
complexes of kinases and their temporal substrates.

Srihari et al. also noticed that by incorporating such temporal information the ``dynamics" of protein complexes could be better understood.
They observed an interesting relationship between the ``staticness" (constitutive expression) of a protein and its participation in 
multiple complexes\cite{Srihari2012poster} - cells tend to maintain generic proteins as `static' to enable
their ``reusability" across multiple temporal complexes.

\subsection{Challenges in detecting membrane complexes}
Membrane protein complexes are formed by physical interactions among
membrane proteins. Membrane proteins are attached to or associated with
the membranes of the cell or its organelles. Membrane proteins constitute
approximately $30\%$ of the proteomes of organisms, yet they are one of the least studied subsets of proteins.
The study of membrane proteins and their complexes is crucial in understanding diseases and aiding new drug
discoveries\cite{vonHeijne2007}.

Membrane protein complexes are notoriously difficult to study using traditional
high-throughput techniques like Y2H and TAP-MS\cite{Lalonde2008}.
This is due in part to the hydrophobic (insoluble) nature of membrane proteins, as well as the ready dissociation of
subunit interactions, either between transmembrane subunits or between
transmembrane and cytoplasmic subunits\cite{Lalonde2008,Barrera2008}.

In order to counter the disadvantages of conventional techniques, new
biochemical techniques have been developed recently to facilitate the
characterization of interactions among membrane proteins. Among these is
the split-ubiquitin membrane yeast two-hybrid (MYTH) system\cite{Miller2005,Kittanakom2009,Petschnigg2011}.
With the development of the high-throughput MYTH system, a fair number of interactions among membrane proteins 
have been recently catalogued in species such as \emph{Arabidopsis thaliana}\cite{Lalonde2008} and yeast 
\emph{Saccharomyces cerevisiae}\cite{Miller2005,Kittanakom2009}.

The identification of membrane complexes requires understanding their
assembly - how the individual proteins come together to form complexes, and
how these complexes are eventually degraded. This is because membrane
proteins are not stable entities as their soluble counterparts.
Studies reveal that this assembly occurs in an orderly fashion, that is,
membrane complexes are formed by an ordered assembly of intermediaries,
and in order to prevent unwanted intermediaries, this assembly is highly aided
by chaperones\cite{Daley2008}. 
Many membrane complexes are formed by transient interactions involving
exchange of proteins in and out of existing complexes via membranes - a phenomenon called
`dynamic exchange'\cite{Daley2008}.
The need therefore now is to develop sophisticated algorithms that take into account these aspects specific to membrane complexes 
to mine them effectively from membrane sub-interactomes.

%%%

\section{Conclusions}
Protein complexes are the fundamental functional units responsible for many
biological mechanisms within the cell. Their identification is therefore necessary
to understand the cellular organization and machinery. The advent of high-throughput techniques for inferring protein interactions
in a large-scale fashion has fueled development of computational techniques to systematically mine for potential complexes
from the network of interactions. In this work, we reviewed, classified and evaluated some of the key computational methods developed
till date for the detection of protein complexes from PPI networks.
We presented two insightful taxonomies of existing methods - `bin-and-stack' and `tree'. From these taxonomies
we note that scoring of raw interaction datasets (followed by filtering of false positives) and integrating key biological insights with
topology can significantly improve complex prediction.

Even though more than 20 methods have been developed over the years, complex detection still requires
careful attention in handling errors and noise in experimental datasets, and reconstructing complexes with high accuracies.
On this front, we identified some of the crucial limitations and challenges facing current 
experimental and computational techniques.
Interaction datasets coming from different experimental sources show surprising lack of correlation and also contain sizeable
fraction of spurious (false positive) interactions. This severely impacts the accuracy and coverage of complex detection methods.
Further, computational methods also overly rely on the assumption that complexes are embedded among densely connected groups of proteins, 
an assumption that is not fully valid in the wake of insufficient credible interactions.
Finally, the interactions among membrane proteins have not been catalogued 
adequately making it difficult to identify an important group of complexes necessary for understanding diseases - membrane complexes.

We hope that our review and assessment of computational methods as well as the challenges highlighted here 
will provide valuable insights to drive future research for further advancing the `state-of-the-art' in
computational prediction, characterization and analysis of protein complexes from organisms.

%%%%%%%%%%%
\section*{Acknowledgments}
This work was supported in part by the National University of Singapore under ARF grants
R-252-000-361-112 and R-252-000-461-112. SS is now associated with the Institute of Molecular Bioscience, University of Queensland,
and would like to thank the institute for providing the resources to complete this work.


\begin{thebibliography}{0}
%\bibitem{beeson} Beeson MJ, {\it Foundations of Constructive Mathematics}, 
%Springer, Berlin, 1985.




\bibitem{Spirin2003}
Spirin, V., Mirny, L., {Protein complexes and functional modules in molecular networks},
\emph{Proc. Natl. Acad. Sci.} {\bf100(21)}:12123--12128, 2003.


\bibitem{Hartwell1999}
Hartwell, L.H., Hopfield, J.J., Leiber, S., Murray, A.W.,
{From molecular to modular cell biology},
\emph{Nature} {\bf 402(6761 Suppl)}:C47--52, 1999.



\bibitem{Zhang2008}
Zhang, B., Park, B.H., Karpinets, T., Samatova, N., {From pull-down data to protein interaction networks and complexes with biological relevance}, 
\emph{Bioinformatics} {\bf 24(7)}:979--986, 2008.




\bibitem{Uetz2000}
Uetz, P., Giot, L., Cagney, G., Traci, A., Judson R., Knight, S.R., Lokshon, D., Narayan, V., Srinivasan, M., Pochart, P., Qureshi-Emili, A., Li, Y., Godwin, B., 
Conover, D., Kalbfleish, T., Vijayadamodar, G., Yang, M., Johnston, M., Fields, S., Rothberg, J.M.,
{A comprehensive analysis of protein-protein interactions in Saccharomyces cerevisiae}, 
\emph{Nature} {\bf 403(6770)}:623--627, 2000.


\bibitem{Ito2001}
Ito T., Chiba, T., Ozawa, R., Yoshida, M., Hattori, M., Sakaki, Y., 
{A comprehensive two-hybrid analysis to explore the yeast protein interactome}, 
\emph{Proc. Natl. Acad. Sci.} {\bf 98(8)}:4569--4574, 2001.



\bibitem{Michinck2003}
Michinck, S.W., {Protein fragment complementation strategies for biochemical network mapping}, 
\emph{Curr. Opin. in Biotech.} {\bf 14(6)}:610-617, 2003.


\bibitem{Gavin2002}
Gavin, A.C., Bosche, M., Krause, R., Grandi, P., Marzioch, M., Bauer, A., Schultz, J., Rick, J.M., Michon, A-M., Cruciat, C-M., Remor, C., Hofert, C., Schelder,M., Brajenovic,M., Ruffner,M., Merino,A., Klein,K., Hudak, M., Dickson,D., Rudi,T., Gnau,V., Bauch,A., Bastuck,S., Huhse,B., Leutwein,C., Heurtier,M-A., Copley,R-R., Edelmann,A., Querfurth,E., Rybin,V., Drewes,G., Raida, M., Bouwmeester, T., Bork,P., Seraphin,B., Kuster,B., Neubauer,G., Superti-Furga,G.,  
{Functional organization of the yeast proteome by systematic analysis of protein complexes},
\emph{Nature} {\bf 415(6868)}:141--147, 2002.


\bibitem{Ho2002}
Ho, Y., Gruhler, A., Heilbut, A., Bader, G., Moore, L., Adams, S.L., Millar, A., Taylor, P., Bennett, K., Boutilier, K., Yang, L., Wolting, C., Donaldson, I., Schandorff, S., Shewnarane, J., Vo, M., Taggart, J., Goudreault, M., Muskat, B., Alfarano, C., Dewar, D., Lin, Z., Michalickova, K., Willems, A.R., Sassi, H., Nielsen, P.A., Rasmussen, K.J., Andersen, J.R., Johansen, L.E., Hansen, L.H., Jespersen, H., Podtelejnikov, A., Nielsen, E., Crawford, J., Poulsen, V., Sorensen, B.D., Matthiesen, J., Hendrickson, R.C., Gleeson, F., Pawson, T., Moran, M.F., Durocher, D., Mann, M., Hogue, C.W., Figeys, D., Tyers, M., 
{Systematic identification of protein complexes in Saccharomyces cerevisiae by mass spectrometry},
\emph{Nature} {\bf 415(6868)}:180--183, 2002.


\bibitem{Gavin2006}
Gavin, A.C., Aloy, P., Grandi, P., Krause, R., Boesche, M., Marzioch, M., Rau, C., Jensen, L.J., Bastuck, S., Dumpelfeld, B., Edelmann, A., Heurtier, M.A., Hoffman, V., Hoefert, C., Klein, K., Hudak, M., Michon, A.M., Schelder, M., Schirle, M., Remor, M., Rudi, T., Hooper, S., Bauer, A., Bouwmeester, T., Casari, G., Drewes, G., Neubauer, G., Rick, J.M., Kuster, B., Bork, P., Russell, R.B., Superti-Furga, G., 
{Proteome survey reveals modularity of the yeast cell machinery},
\emph{Nature} {\bf 440(7084)}:631--636, 2006.


\bibitem{Krogan2006}
Krogan, N.J., Cagney, G., Yu, H., Zhong, G., Guo, X., Ignatchenko, A., Li, J., Pu, S., Datta, N., Tikuisis, A.P., Punna, T., Peregrin-Alvarez, J.M., Shales, M., Zhang, X., Davey, M., Robinson, M.D., Paccanaro, A., Bray, J.E., Sheung, A., Beattie, B., Richards, D.P., Canadien, V., Lalev, A., Mena, F., Wong, P., Starostine, A., Canete, M.M., Vlasblom, J., Wu, S., Orsi, C., Collins, S.R., Chandran, S., Haw, R., Rilstone, J.J., Gandi, K., Thompson, N.J., Musso, G., Onge, P., Ghanny, S., Lam, M.H., Butland, G., Altaf-Ul, A.M., Kanaya, S., Shilatifard, A., O'Shea, E., Weissman, J.S., Ingles, C.J., Hughes, T.R., Parkinson, J., Gerstein, M., Wodak, S.J., Emili, A., Greenblatt, J.F.,
{Global landscape of protein complexes in the yeast Saccharomyces cerevisiae},
\emph{Nature} {\bf 440(7084)}:637--643, 2006.


\bibitem{vonMering2002}
von Mering, C., Krause, R., Snel, B., Cornell, M., Oliver, S.G., Fields, S., Bork, P., 
{Comparative assessment of large-scale datasets of protein-protein interactions},
\emph{Nature} {\bf 417(6887)}:399--403, 2002.


\bibitem{Bader2002}
Bader, G.D., Hogue, C.W.V., 
{Analyzing yeast protein-protein interaction data obtained from different sources},
\emph{Nature Biotechnology} {\bf 20(10)}:991--997, 2002.


\bibitem{Cusick2008}
Cusick, M.E., Yu, H., Smolyar, A., Venkatesan, K., Carvunis, A-R., Simonis, N., Rual, J-F., Borick, H.,
Braun, P., Dreze, M., Vandenhaute, J., Galli, M., Yazaki, J., Hill, D.E., Ecker, J.R., Roth, F.P., Vidal, M.,
{Literature-curated protein interaction datasets},
\emph{Nature Methods} {\bf 6(1)}:39--46, 2009.


\bibitem{Mackay2008}
Mackay, J.P., Sunde, M., Lowry, S., Crossley, M., Matthews, J.M., 
{Protein interactions to believe or not to believe?},
\emph{Trends in Biochemical Sciences} {\bf 30(1)}:242--243, 2008.




%% Scoring schemes

\bibitem{Collins2007}
Collins, S.R., Kemmeren P., Zhao, X.C., Greenbalt, J.F., Spencer F., Holstege, F., Weissman, J.S., Krogan, N.J.,
{Toward a comprehensive atlas of the physical interactome of Saccharomyces cerevisiae},
\emph{Molecular Cellular Proteomics} {\bf 6(3)}:439--450, 2007.




\bibitem{Hart2007}
Hart, G., Lee, I., Marcotte, E.R., 
{A high-accuracy consensus map of yeast protein complexes reveals modular nature of gene essentiality},
\emph{BMC Bioinformatics} {\bf 8}:236, 2007.



\bibitem{Chua2008}
Chua, H.N., Sung, W.K., Wong, L.,
{Exploiting indirect neighbours and topological weight to predict protein function from protein protein interactions},
\emph{Bioinformatics} {\bf 22(13)}:1623--1630, 2006.


\bibitem{Liu2008}
Liu, G., Li, J., Wong, L., 
{Assessing and predicting protein interactions using both local and global network topological metrics},
\emph{Genome Informatics Series: Proceedings of the 19th International Conference on Genome Informatics} 
{\bf 21}:138--149, 2008.


\bibitem{Friedel2008}
Friedel C., Krumsiek, J., Zimmer, R., 
{Bootstrapping the interactome: unsupervised identification of protein complexes in yeast},
\emph{J. Computational Biology} {\bf 16(8)}:971--987, 2009.



\bibitem{Kuchaiev2009}
Kuchaiev, O., Rasajski, M., Higham, D., Przulj, N., 
{Geometric de-noising of protein-protein interaction networks},
\emph{PLoS Computational Biology} {\bf 5(8)}:e1000454, 2009.


\bibitem{Voevodski2009}
Voevodski, K., Teng, S-H., Yu, X., 
{Spectral affinity in protein networks},
\emph{BMC Systems Biology} {\bf 3}:112, 2009.


\bibitem{Jain2010}
Jain, S., Bader, G., 
{An improved method for scoring protein-protein interactions using semantic similarity within the gene ontology},
\emph{BMC Bioinformatics} {\bf 11}:562, 2010.


\bibitem{Suthram2006}
Suthram, S., Shlomi, T., Ruppin, E., Sharan, R., Ideker, T., 
{A direct comparison of protein interaction confidence assignment schemes},
\emph{BMC Bioinformatics} {\bf 7}:360, 2006.


\bibitem{SrihariThesis2012}
Srihari, S., Integrating Biological Insights with Topological Characteristics for Improved Complex Prediction from Protein Interaction Networks,
PhD Thesis, National University of Singapore, 2012.


\bibitem{Brohee2006}
Brohee, S., van Helden, J., 
{Evaluation of clustering algorithms for protein-protein interaction networks},
\emph{BMC Bioinformatics} {\bf 7}:488, 2006.


\bibitem{Vlasblom2009}
Vlasblom, J., Wodak, S., 
{Markov clustering versus affinity propagation for the partitioning of protein interaction graphs},
\emph{BMC Bioinformatics} {\bf 10}:99, 2009.


\bibitem{Xli2010}
Li, X.L., Wu, M., Kwoh, C.C., Ng, S.K.,
{Computational approaches for detecting protein complexes from protein interaction networks: a survey},
\emph{BMC Genomics} {\bf 11(Suppl 1)}:S3, 2010.


%%%%%%%%







\bibitem{Kalna2007}
Kalna, G., Higham, D., 
{A clustering coefficient for weighted networks, with application to gene expression data},
\emph{J. AI Communications - Network Analysis in Natural Sciences and Engineering} {\bf 20(4)}:263--271, 2007.




\bibitem{Bader2003}
Bader, G.D., Hogue, C.W.V., 
{An automated method for finding molecular complexes in large protein interaction networks},
\emph{BMC Bioinformatics} {\bf 4}:2, 2003.



\bibitem{Dongen2000}
van Dongen S., {Graph clustering by flow simulation},
PhD thesis, University of Utrecht, 2000.



\bibitem{Enright2002}
Enright, A.J., van Dongen, S., Ouzounis, C.A., 
{An efficient algorithm for large-scale detection of protein families},
\emph{Nucleic Acids Research} {\bf 30(7)}:1575--1584, 2002.


\bibitem{Pu2007}
Pu, S., Vlasblom, J., Emili, A., Greenbalt, J., Wodak, S.J., 
{Identifying functional modules in the physical interactome of Saccharomyces cerevisiae},
\emph{Proteomics} {\bf 7(6)}:944--960, 2007.


\bibitem{Enright2004}
Pereira-Leal, J.B., Enright, A.J., Ouzounis, C.A.,
{Detection of functional modules from protein interaction networks},
\emph{Proteins} {\bf 54(1)}:49--57, 2004.



\bibitem{Liu2009}
Liu, G., Wong, L., Chua, H.N., 
{Complex discovery from weighted PPI networks},
\emph{Bioinformatics} {\bf 25(15)}:1891--1897, 2009.




\bibitem{Adamcsek2003}
Adamcsek, B., Palla, G., Farkas, I., Derenyi, I., Vicsek, T., 
{CFinder: locating cliques and overlapping modules in biological networks},
\emph{Bioinformatics} {\bf 22(8)}:1021--1023, 2006.



\bibitem{Li2005}
Li X.L., Tan, S.H., Foo, C.S., Ng, S.K., 
{Interaction Graph mining for protein complexes using local clique merging},
\emph{Genome Informatics Series: Proceedings of the 16th International Conference on Genome Informatics} {\bf 16(2)}:260--269, 2005.



\bibitem{Tomita2006}
Tomita, E., Tanaka, A., Takahashi, H., 
{The worst-case time complexity for generating all maximal cliques and computational experiments},
\emph{J. Theoretical Computer Science} {\bf 363(1)}:28--42, 2006.



\bibitem{Nepusz2012}
Nepusz, T., Yu, H., Paccanaro, A., 
{Detecting overlapping protein complexes in protein-protein interaction networks},
\emph{Nature Methods} {\bf 9}:471--472, 2012.



\bibitem{Chua2007}
Chua H.N., Ning, K., Sung, W.K., Leong, H.W., Wong, L., 
{Using indirect protein-protein interactions for protein complex prediction},
\emph{J. Bioinformatics Computational Biology} {\bf 6(3)}:435--466, 2008.




\bibitem{Wang2009}
Wang, H., Kakaradov B., Collins S.R., Karotki, L., Fiedler, D., Shales M., Shokat, K.M., Walter, T., Krogan N.J., Koller, D., 
{A complex-based reconstruction of the Saccharomyces cerevisiae interactome},
\emph{Molecular Cellular Proteomics} {\bf 8}:1361--1377, 2009.


\bibitem{Zhang2011}
Zhang, J., {A dynamical method to extract communities induced by low or middle-degree nodes},
\emph{Proceedings of the IEEE International Conference on Systems Biology (ISB)}, pp.340--344, 2011.


\bibitem{Ma2011}
Ma, X., Gao, L., {Detecting protein complexes in PPI network roles of interactions},
\emph{Proceedings of the IEEE International Conference on Systems Biology (ISB)}, pp.223--239, 2011.


\bibitem{Wang2011}
Wang, Y., Gao, L., Chen, Z., {An edge based core-attachment method to detect protein complexes in PPI networks},
\emph{Proceedings of the IEEE International Conference on Systems Biology (ISB)} pp.72--77, 2011.


\bibitem{Chin2010}
Chin, C.H., Chen, S.H., Ho, C.W., Ko, M.T., Lin, C.Y., 
{A hub-attachment based method to detect functional modules from confidence-scored protein interactions and expression profiles},
\emph{BMC Bioinformatics} {\bf 11(Suppl 1)}:S25, 2010.




\bibitem{WuMin2009}
Wu, M., Li, X., Ng S.K., 
{A core-attachment based method to detect protein complexes in PPI networks},
\emph{BMC Bioinformatics} {\bf 10}:169, 2009.



\bibitem{Leung2009}
Leung, H., Xiang, Q., Yiu, S.M., Chin, F.Y., 
{Predicting protein complexes from PPI data: a core-attachment approach},
\emph{J. of Computational Biology} {\bf 16(2)}:133--144, 2009.


\bibitem{Srihari2009}
Srihari, S., Ning, K., Leong, H.W., 
{Refining Markov Clustering for complex detection by incorporating core-attachment structure},
\emph{Genome Informatics Series: Proceedings of the 20th International Conference on Genome Informatics} {\bf 23(1)}:159--168, 2009.



\bibitem{Srihari2010}
Srihari, S., Ning, K., Leong, H.W., 
{MCL-CAw: a refinement of MCL for detecting yeast complexes from weighted PPI networks by incorporating core-attachment structure},
\emph{BMC Bioinformatics} {\bf 11}:504, 2010.






\bibitem{Ashburner2000}
Ashburner, M., Ball C.A., Blake J.A., Botstein, D., Butler, H., Cherry, M., Davis AP, Dolinski K, Dwight SS, Eppig JT, Harris MA, Hill DP, Issel-Tarver L, Kasarskis A, Lewis S, Matese JC, Richardson JE, Ringwald M, Rubin GM, Sherlock G.,
{Gene ontology: a tool for the unification of biology},
\emph{Nature Genetics} {\bf 25(1)}:25--29, 2000.



\bibitem{King2004}
King, A.D., Przulj, N., Jurisca, I., 
{Protein complex prediction via cost-based clustering},
\emph{Bioinformatics} {\bf 20(17)}:3013--3020, 2004.




\bibitem{Li2007}
Li, X.L., Foo C.S., Ng, S.K., 
{Discovering protein complexes in dense reliable neighborhoods of protein interaction networks},
\emph{Computational Systems Bioinformatics Series: Proceedings of the Computational Systems Bioinformatics Conference} {\bf 6}:157--168, 2007.


\bibitem{Sharan2004}
Sharan, R., Ideker, T., Kelley, B.P., Shamir, R., Karp, R.M., 
{Identification of protein complexes by comparative analysis of yeast and bacterial protein interaction data},
\emph{J. Computational Biology} {\bf 12(6)}:835--846, 2005.


\bibitem{Sharan2006}
Hirsh, R., Sharan, R., 
{Identification of conserved protein complexes based on a model of protein network evolution},
\emph{Bioinformatics} {\bf 23(2)}:e170--e176, 2006.


\bibitem{Sharan2007}
Dost, B., Shlomi, T., Gupta, N., Ruppin, E., Bafna, V., Sharan, R., 
{QNet: A Tool for Querying Protein Interaction Networks},
\emph{J. Computational Biology} {\bf 15(7)}:913--925, 2008.


\bibitem{Kim2006}
Kim, P.M., Lu, L.J., Xia, Y., Gerstein, M.B., 
{Relating three-dimensional structures to protein networks provides evolutionary insights},
\emph{Science} {\bf 314(5807)}:1938--1941, 2006.


\bibitem{Ozawa2010}
Ozawa, Y., Saito, R., Fujimori, S., Kashima, H., Ishizaka, M., Yanagawa, H., Miyamoto-Sato, E., Tomita, M.,
{Protein complex prediction via verifying and reconstructing the topology of domain-domain interactions},
\emph{BMC Bioinformatics} {\bf 11}:350, 2010.


\bibitem{Jung2010}
Jung, S.H., Hyun, B., Jang, W.H., Hur, H.Y., Han, D.S., 
{Protein complex prediction based on simultaneous protein interaction network},
\emph{Bioinformatics} {\bf 26(3)}:385--391, 2009.


\bibitem{Przytycka2010}
Przytycka, T., Singh, M., Slonim, D.K., 
{Toward the dynamic interactome: it's about time},
\emph{Briefings in Bioinformatics} {\bf 11(1)}:15--29, 2010.




\bibitem{Blatt1996}
Blatt, M., Wiseman, S., Domany, E., 
{Superparamagnetic clustering of data},
\emph{Physical Review Letters} {\bf 76(18)}:3251--3254, 1996.



\bibitem{Mewes2006}
Mewes, H.W., Amid, C., Arnold, R., Frishman, D.  
{MIPS: analysis and annotation of proteins from whole genomes},
\emph{Nucleic Acids Research} {\bf 32(Suppl 1)}:D41--D44, 2004.


\bibitem{Frey2007}
Frey, B.J., Dueck, D., 
{Clustering by passing messages between data points},
\emph{Science} {\bf 315(5814)}:972--976, 2007.



\bibitem{Xenarios2002}
Xenarious, I., Salwinski, L., Duan, J.X., Higney, P., Kim, S-L., Eisenberg, D., 
{DIP, the Database of Interacting Proteins a research tool for studying cellular networks of protein interactions},
\emph{Nucleic Acids Research} {\bf 30(1)}:303--305, 2002.


\bibitem{Pu2009}
Pu, S., Wong, J., Turner, B., Cho, E., Wodak, S., 
{Up-to-date catalogues of yeast protein complexes},
\emph{Nucleic Acids Research} {\bf 37(3)}:825-831, 2009.



\bibitem{Aloy2004}
Aloy, P., Bottcher, B., Ceulemans, H., Leutwein, C. Mellwig, C., Fischer, S., Gavin, A.C., Bork, P., Superti-Furga, G., Serrano, L., Russell, R.,
{Structure-based assembly of protein complexes of yeast},
\emph{Science} {\bf 303(5666)}:2026--2029, 2004.


\bibitem{Cherry98}
Cherry, J.M., Adler, C., Chervitz S.A., Dwight S.S., Jia, Y., Juvik, G., Roe, T., Schroeder, M., Weng, S., Borstein, D.,
{SGD: Saccharomyces Genome Database},
\emph{Nucleic Acids Research} {\bf 26(1)}:73--79, 1998.


\bibitem{Srihari2012}
Srihari, S., Leong, H.W., 
{Employing functional interactions for the characterization and detection of sparse complexes from yeast PPI networks},
to appear in
\emph{International J. Bioinformatics Research and Applications}, 2012.


\bibitem{Srihari2012poster}
Srihari, S., Leong, H.W., 
{Temporal Dynamics of Protein Complexes in PPI Networks: A Case Study using Yeast Cell Cycle Dynamics},
to appear in \emph{BMC Bioinformatics}, 2012.


\bibitem{vonHeijne2007}
von Heijne, G., 
{The membrane protein universe whats out there and why bother?},
\emph{J. Internal Medicine} {\bf 261(6)}:543--557, 2007.


\bibitem{Lalonde2008}
Lalonde, S., Ehrhardt, D.W., Logue, D., Chen, J., Rhee, S.Y., Frommer, W.B., 
{Molecular and cellular approaches for the detection of protein-protein interactions latest techniques and current limitations},
\emph{Plant J.} {\bf 53(4)}:610-635, 2008.


\bibitem{Barrera2008}
Barrera, N.P., Bartolo, N.D., Booth, P.J., Robinson, C.V., 
{Micelles Protect Membrane Complexes from Solution to Vacuum}, 
\emph{Science} {\bf 321(5886)}:243--246, 2008.


\bibitem{Miller2005}
Miller, J.P., Russel.S. Ben-Hur, A., Desmarais, C., Stagljar, I., Novel, W.S., Fields, S., 
{Large-scale identification of yeast integral membrane protein interactions}, 
\emph{Proc. Natl. Acad. Sci.} {\bf 102(34)}:12123-12128, 2005.



\bibitem{Kittanakom2009}
Kittanakom, S., Chuk, M., Wong, V., Snyder, J., Edmonds, D., Lydakis, A., Zhang, Z., Auerbach, D., Stagljar, I., 
{Analysis of membrane protein complexes using the split-ubiquitin membrane yeast two-hybrid (MYTH) system}, 
\emph{Methods Mol. Bio.} {\bf 548}:247--271, 2009.


\bibitem{Petschnigg2011}
Petshnigg, J., Moe, O.W., Stagljar, I., 
{Using yeast as a model to study membrane proteins},
\emph{Curr. Opin. in Nephr. Hypertension} {\bf 20(4)}:425--432, 2011.


\bibitem{Daley2008}
Daley, D.O., 
{The assembly of membrane proteins into complexes}.
\emph{Curr. Opin. Struct. Biol.} {\bf 8(4)}:420--424, 2008.


\end{thebibliography}
\end{document}